\documentclass[a4paper,fleqn,usenatbib,useAMS]{mnras}

\usepackage{graphicx}	
\usepackage{amsmath}	
\usepackage{amssymb}	
\usepackage{multicol}        
\usepackage{bm}		
\usepackage{multirow}
\usepackage{booktabs,caption}
\usepackage[flushleft]{threeparttable}

\usepackage[T1]{fontenc}
\usepackage{ae,aecompl}

\usepackage{newtxtext,newtxmath}

\title[A Search for Supernova Light Echoes in NGC 6946]{A Search for Supernova Light Echoes in NGC 6946 with SITELLE}

\author[Radica et al.]{M. C. Radica,$^{1}$\thanks{Current address: D\'epartement de physique, Universit\'e de Montr\'eal, 1375 Avenue Th\'er\`ese-Lavoie-Roux, Montr\'eal, QC, H2V 0B3, Canada}\thanks{E-mail: \href{mailto:radica@astro.umontreal.ca}{radica@astro.umontreal.ca}} D. L. Welch,$^{1}$ and L. Rousseau-Nepton$^{2,3}$
\\
$^{1}$Department of Physics and Astronomy, McMaster University, 1280 Main St W, Hamilton, ON, L8S 4L8, Canada\\
$^{2}$Canada-France-Hawaii Telescope, 65-1238 Mamalahoa Hwy, Kamuela, HI, 96743, USA\\
$^{3}$Department of Physics and Astronomy, University of Hawaii at Hilo, 200 W Kawili St, Hilo, HI, 96720, USA}

\date{Last updated 2015 May 22; in original form 2013 September 5}

\pubyear{2020}

\begin{document}
\label{firstpage}
\pagerange{\pageref{firstpage}--\pageref{lastpage}}
\maketitle

\begin{abstract}
We present the analysis of four hours of spectroscopic observations of NGC 6946 with the SITELLE Imaging Fourier Transform Spectrometer on the Canada-France-Hawaii Telescope, acquired to search for supernova light echoes from its ten modern supernovae. We develop a novel spectroscopic search method: identifying negatively sloped continua in the narrow-band SN3 filter as candidate highly-broadened P-Cygni profiles in the H$\alpha$ line, which would be characteristic of the spectra of supernovae ejecta. We test our methodology by looking for light echoes from any of the ten supernovae observed in NGC 6946 in the past 100 years. We find no evidence of light echoes above the survey surface brightness limit of 1$\times$10$^{-15}$\,erg/s/cm$^2$/arcsec$^2$.
\end{abstract}

\begin{keywords}
galaxies: individual: NGC 6946 - supernovae: general - instrumentation: SITELLE - techniques: imaging spectroscopy
\end{keywords}


\section{Introduction}
\label{sec:Introduction} 

\subsection{Supernova Light Echoes}

The timescales over which supernovae (SNe) evolve are unusually short compared to many astronomical objects -- they change dramatically on human and historical timescales. Supernova (SN) light echoes (LEs) are formed when some small fraction of the light from a SN is scattered by dust into the line of sight of the observer. They are a unique way to study the properties of a SN outburst long after direct light from the event has passed the observer. \citet{1940RvMP...12...66Z} was the first to suggest that SNe could be studied through observations of their LEs, and the first LE surveys were attempted by \citet{1965PASP...77..269V}, and  \citet{1966PASP...78...74V}.

Since LEs were first formally understood in 1939 \citep{couderc1939aureoles}, they have proved powerful methods for late-time study of transient events. The scattered (optical) light preserves both the spectral line properties, and a convolved form of the light curve of the initial SN. These can be used for ``posthumous'' classification of ancient SNe for which no modern observations exist as was demonstrated in the case of Tycho Brahe's 1572 SN, identified as a Type Ia in 2008 through observations of its LEs \citep{krause2008tycho, rest2008scattered}.

Systems also exist in the Milky Way (MW) and Large Magellanic Cloud (LMC) for which multiple LE systems have been discovered. Each individual LE represents a unique viewing angle to the source event, allowing for a degree of three-dimensional reconstruction of photospheric properties -- one of the few areas in astronomy where this is possible. The usefulness of these multiple perspectives in directly confirming asymmetry has been demonstrated for the Cassiopeia A and SN 1987A outbursts \citep{rest2011direct, sinnott2013asymmetry}.

LEs are increasingly being employed in the literature to account for the late-time evolution of SN light curves of both Type Ia, and Type II events (e.g. \citealp{milne2019slowly, maund2019origin}). Direct observational detections of SN LEs though are still relatively rare -- one of the most fruitful target being the LMC, where LE searches are still underway (e.g. \citealp{margheim2019light}). The luminous blue variable (LBV) star $\eta$ Carinae is also being actively studied through its multiple LEs, especially from the epoch of its Great Eruption, with indications that ejecta speeds likely exceeded 10,000\,km/s \citep{ smith2018exceptionally, smith2018light}. 

Scattered LEs are much fainter than the brightest portion of the original SN lightcurve -- compare a peak apparent magnitude of $+2.9$ for SN 1987A, to its brightest LEs which had a surface brightness of +19.3\,mag/arcsec$^{2}$ \citep{west1987astrometry, xu1994two, sinnott2013asymmetry}. In addition, the likelihood of detecting a LE is confounded by large uncertainties in dust distribution, age, distance, and foreground extinction. For Type II SNe there are additional uncertainties stemming from the wider range in outburst luminosities, and dominantly lower peak brightnesses when compared to SNe of Type Ia \citep{young1989absolute, phillips1993absolute}. It is no surprise then, that LEs from Type II SNe are under-represented in current MW and LMC surveys. Considering the relative rarity of SNe, one per 50 years in the average galaxy and no naked-eye SNe since 1604 in the MW, the duration of LE detectability with age is poorly constrained \citep{cappellaro1999new, diehl2006radioactive}. An object containing multiple Type II SNe with known ages holds out the prospect of providing a firmer basis for our statistics of LE detectability - particularly for Type II events.

\subsection{NGC 6946}

\begin{figure}
	\centering
	\includegraphics[width=\columnwidth]{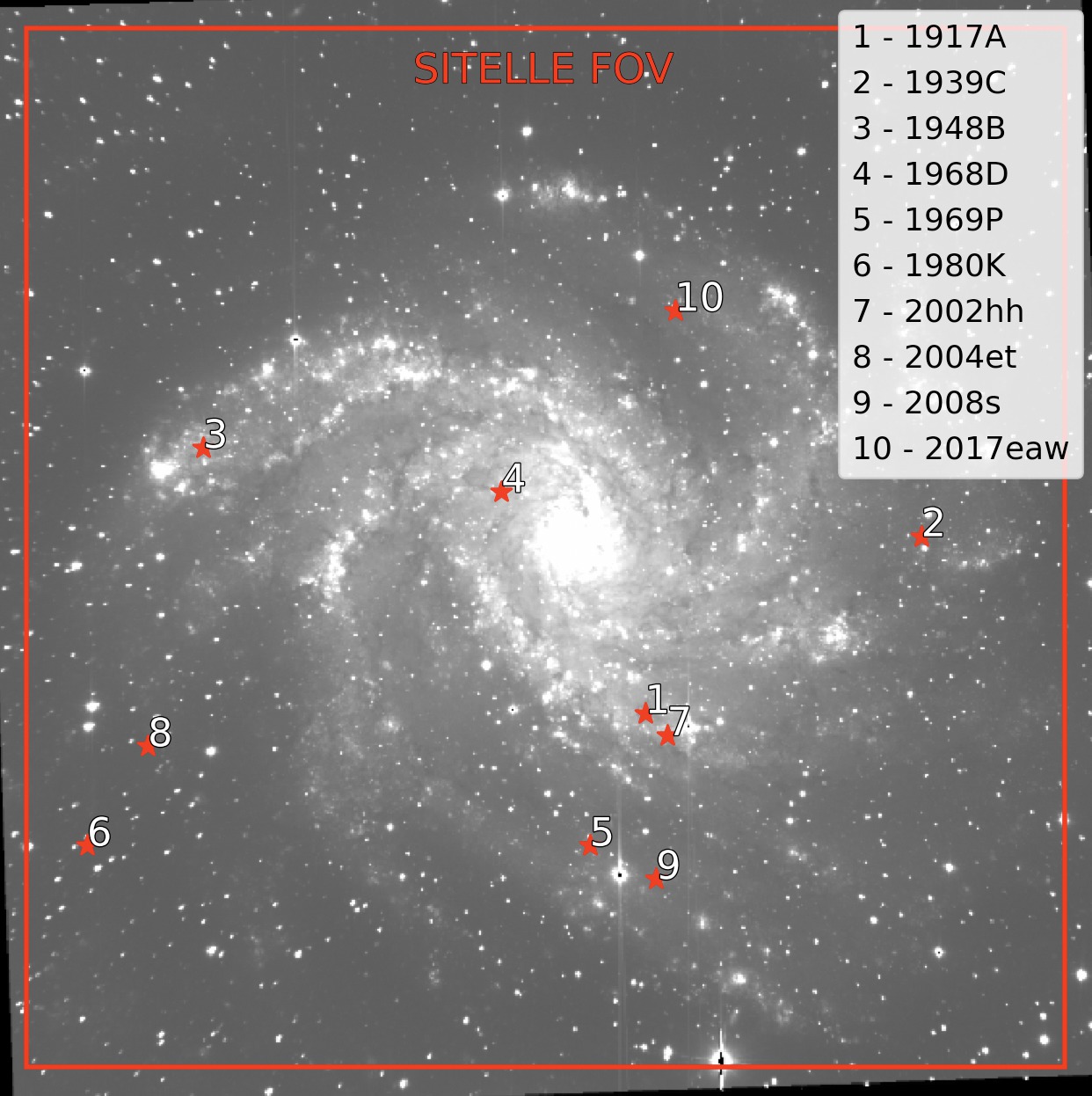}
    \caption{Locations of the ten SNe within NGC 6946. The red star symbols represent the location of each SN. The background image is the deep frame from our SITELLE observations. NGC 6946 fits well within the 11\arcmin$\times$11\arcmin$\,$ FOV of SITELLE (red square).}
    \label{fig:SNe_pos}
\end{figure}

The nearby, face-on spiral galaxy NGC 6946, at a distance of 7.72\,Mpc, is such an object \citep{anand2018robust}. Remarkably, NGC 6946 has had \textit{ten} SNe discovered in it during the past 100 years -- the first recorded in 1917 and the most recent in 2017. The galaxy therefore could contain a set of potentially detectable LEs from SN events at various stages of decay \citep{sugerman2012thirty}. It would seem to be the ideal object in which to undertake a study of the detectability of LEs from Type II SNe. The locations of the ten SNe within the galaxy are shown in Figure \ref{fig:SNe_pos}.

The ten SNe in NGC 6946 are relatively evenly distributed over the 100-year period. The events in the later part of the 20$\rm ^{th}$ and early 21$\rm ^{st}$ century are well-classified and have many photometric and spectroscopic observations subsequent to the initial discovery across multiple wavebands. However, many of the earliest SNe have no proper classification due to a lack of timely spectra. The classification of one of the SNe, SN 2008S, is currently debated in the literature due to its low outburst luminosity. The discovery of a low mass, dust clouded progenitor \citep{prieto2008discovery} has fueled debate over whether SN 2008S was a Type IIn event as initially posited, or the eruption of a LBV star (e.g. \citealp{thompson2009new, kochanek2011dusty}). The main astrometric, spectroscopic, and photometric properties of the SNe are summarized in Table \ref{tab:asproperties}.

\begin{table*}
\centering
\begin{threeparttable}
\caption{Summary of the major properties of the ten SNe in NGC 6946.}  
\begin{tabular}{c|ccccccc}
\hline
\hline
\multirow{2}{*}{Designation} & \multirow{2}{*}{Discovery Date} & \multirow{2}{*}{RA} & \multirow{2}{*}{Dec} & \multirow{2}{*}{Offset} & \multirow{2}{*}{Type} & Discovery  & \multirow{2}{*}{References} \\
 & & & & & & Apparent Magnitude & \\
\hline
1917A & 19 July, 1917 & 20:34:46.9 & +60:07:29 & 37"W, 105"S & II $^*$ & +14.6 & 1,2\\
1939C & 17 July, 1939 & 20:34:24.06 & +60:09:29.9 & 215"W, 24"N & I $^*$ & +13.1 & 2 \\
1948B & 06 July, 1948 & 20:35:21.5 & +60:10:16 & 222"E, 60"N & IIP & +15.1 & 2, 3 \\
1968D & 29 February, 1968 & 20:34:58.40 & +60:09:34.4 & 45.3"E, 19.8"N & II $^*$ &  +13.5 & 2,4 \\
1969P & 11 December, 1969 & 20:34:51.3 & +60:06:14 & 5"W, 180"S & - $^*$ & +13.9 & 2,5 \\
1980K & 28 October, 1980 & 20:35:30.13 & +60:06:23.51 & 280"E, 166"S & IIL & +13.0 & 2,6 \\
2002hh & 31 October, 2002 & 20:34:44.29 & +60:07:19.0 & 60.9"W, 114.1"S & IIP & +16.5 & 7,8 \\
2004et & 27 September, 2004 & 20:35:25.33 & +60:07:17.7 & 247.1"E, 115.4"S & IIP & +12.8 & 9 \\
2008S & 01 February, 2008 & 20:34:45.33 & +60:05:58.4 & 53"W, 196"S & IIn/LBV $^{**}$ & +17.6 & 10 \\
2017eaw & 14 May, 2017 & 20:34:44.238 &  +60:11:36.00 & 61.0"W, 143.0"N & IIP & +12.8 & 11 \\
\hline
\label{tab:asproperties}
\end{tabular}
\begin{tablenotes}
\small
\item $^*$ Insufficient followup for proper classification.
\item $^{**}$ Classification currently debated in the literature (see discussion in text).

$^1$\citet{ritchey1917another} $^2$\citet{schlegel1994x} $^3$\citet{mayall1948supernova} $^4$\citet{hyman1995radio} $^5$\citet{rosino1971possible} $^6$\citet{wood1980supernova} $^7$\citet{stockdale2002supernova} $^8$\citet{faran2014photometric} $^9$\citet{li2005progenitor} $^{10}$\citet{prieto2008discovery} $^{11}$\citet{tsvetkov2018light}

\end{tablenotes}
\end{threeparttable}
\end{table*} 

The ten SNe span a wide range of peak apparent magnitudes, as well as most subclasses of Type II SNe, however with the exception of SN 1948B, all SNe prior to 1980 lacked the sufficient photometric and spectroscopic followup required for direct classification. LEs from any of these SNe would provide the requisite spectra for sub-classification. SN 1969P is particularly noteworthy in that there was no optical followup whatsoever on the SN discovery until almost a year later \citep{rosino1971possible}.

\subsubsection{Light Echoes in NGC 6946}
\label{sec:LEsinNGC}
 
 NGC 6946 has long been identified as potentially housing multiple LE systems. Early work by \citet{boffi1999search} aimed to identify LE candidates based on the optical colours of patches of emission near known SNe. All SNe prior to 2002hh in NGC 6946 were part of their sample, however none of the emission patches they identified yielded promising LE candidates. 

Observations of the mid-infrared (MIR) continuum from SN 2004et strongly suggest the presence of dust along the line-of-sight \citep{sahu2006photometric, misra2007type}. Studying the evolution of the SED over time, \citet{kotak2009dust} found that the late-time SED is best fitted with a three-component model: hot emission from optically-thick gas, warm emission from radiatively-heated dust in the ejecta, and a cold component due to an MIR LE in the interstellar-medium (ISM) dust. SED fitting at 464 days after outburst gives a LE with peak surface brightness of 1.2$\times$10$^{-26}$\,erg/s/cm$^2$/arcsec$^2$ at 50\,\micron. There has been no published follow-up to confirm this. It should also be noted that MIR echoes are caused by dust re-emission, and do not preserve the light curve or spectral progression (with age) of the original SN.

SN 2002hh was directly observed to be coincident with a massive dust cloud in followup Spitzer IRAC observations in all four wavebands, during the epoch from 590-994 days after outburst \citep{barlow2005detection, meikle2006spitzer}. Like SN 2004et, SED fitting indicated the presence of a MIR LE from the SN. Very late-time photometry and spectroscopy by \citet{welch2007extremely} with Gemini/GMOS-N, followed the evolution of the SN from 661 to 1358 days after outburst. No significant fading of the H$\alpha$ flux, nor of flux in the $R$ or $I$ bands was found. Hubble Space Telescope observations confirmed the presence of an extremely bright LE with a peak H$\alpha$ flux of 4$\times$10$^{-13}$\,erg/s/cm$^{2}$/\textrm{\AA}. Eleven years after outburst, the peak H$\alpha$ flux of the LE was reported by \citet{andrews2015late} to have dropped to 3$\times$10$^{-17}$\,erg/s/cm$^{2}/\textrm{\AA}$.

To have the greatest opportunity of recovering either of these LE complexes, as well as potential others from the remaining eight SNe, an observational program was designed which would probe deep into the faint nebular emission of NGC 6946. We outline the observing plan and resulting spectral cube in \S\ref{sec:Observations}, and the various data analyses in \S\ref{sec:Data Analysis}. The results of our survey will be presented and discussed in \S\ref{sec:Results}. Concluding remarks will be made in \S\ref{sec:Conclusions}.


\section{Observations}
\label{sec:Observations}

All previous LE search strategies shared a common limitation -- the selection of fields to maximize the probability of a successful detection. Traditional spectroscopic studies in particular are very much limited by their use of slit spectroscopy. Spectroscopy is usually only performed once LE candidates have been located through difference imaging. For these reasons, spatially-resolved integral field unit (IFU) spectroscopy has the potential to revolutionize the manner in which LE searches are conducted.

SITELLE\footnote{Spectrom\`{e}tre Imageur \`{a} Transform\'{e}e de Fourier pour l'Etude en Long et en Large de raies d'Emission} is an optical imaging Fourier transform spectrometer (IFTS) on the Canada-France-Hawaii Telescope (CFHT). At the time of its commissioning, SITELLE had the largest field of view (FOV) of any optical IFU at 11\arcmin$\times$11\arcmin \citep{drissen2010sitelle}. It has a pixel scale of 0.32\arcsec, and is capable of reaching a spectral resolution R$\sim$10$^4$. SITELLE has six built-in filters covering optical wavebands from 3500 to 9000\,\AA.

With its relatively high and tunable spectral resolution and small pixel scale, SITELLE's primary scientific purpose is spatially-resolved spectroscopy of extragalactic emission line regions \citep{martin2017sitelle, rousseau2018ngc628}. However, SITELLE also possesses attributes which could make it a useful instrument for SN LE searches. As shown in Figure \ref{fig:SNe_pos}, the FOV is large enough to enclose the entire disk of NGC 6946, and the positions of all ten SNe can be captured within in a single pointing. It must be noted that degradation in the image quality was found during commissioning in 2016 -- especially towards the top of the frame, affecting the data presented here \citep{baril2016commissioning}. These issues have now been largely resolved by including one focal plane field corrector at the end of each camera's assembly (Baril et al. 2020 in prep).

We designed an observing program for the 2018B semester (proposal and observation ID 18BC17) in order to use SITELLE to search for LEs from the ten SNe discovered since 1917, as well as from any (unknown) historical SNe for which no outburst was recorded.

\begin{figure} 
	\centering
	\includegraphics[width=\columnwidth]{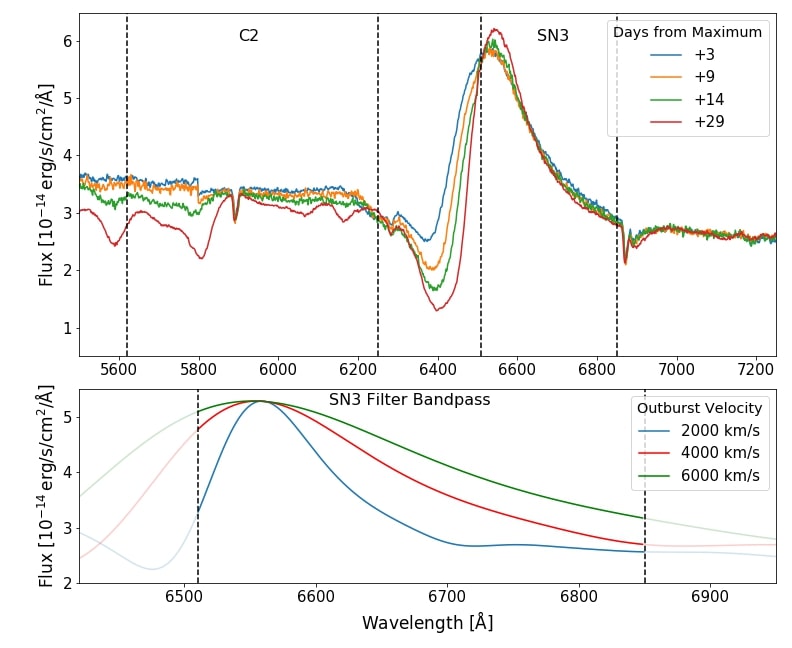}
    \caption{\emph{Top:} Four spectra of SN 2004et spanning a month after initial outburst \citep{sahu2006photometric}. The H$\alpha$ lines display the characteristic P-Cygni profile. The bandpass of the SN3 filter, as well as the neighbouring C2 filter are denoted with dashed black lines.
    \emph{Bottom:} Three P-Cygni profiles for events of different outburst velocities within the SN3 filter bandpass. A P-Cygni profile in the H$\rm \alpha$ line broadened by $\gtrsim$4000\,km/s is indistinguishable from a negative slope in the SN3 filter.}
    \label{fig:broad}
\end{figure}

Due to the Doppler broadening of the H$\alpha$ P-Cygni line profiles, a resolution of only R\,=\,300 would be sufficient for a detection in the dedicated H$\alpha$ filter, SN3 (6510\,-\,6850\,\AA). Figure \ref{fig:broad}, shows the H$\alpha$ P-Cygni line profiles typical of SN events. The bottom panel illustrates that P-Cygni profiles in the H$\rm \alpha$ lines broadened by $\gtrsim$$\rm 4000\,km/s$ would completely fill the SN3 filter bandpass. We therefore did not require the full spectral resolution capabilities of SITELLE, and chose instead to sacrifice spectral resolution in order to obtain a deep image, and potentially detect signatures of the faint LEs.

This tradeoff of spectral resolution for image depth was a key difference of our use of this instrument. Most of the scientific work done with SITELLE uses a few hundred interferometer steps, and short (tens of seconds) exposures per step. We proposed 12 hours of observations with only 49 interferometer steps, but an exposure time of 279 seconds per step. This would allow us to detect H$\alpha$ lines broadened by 200\,\AA\, down to a limit of $\rm 3\times10^{-16}\,erg/s/cm^2/\textrm{\AA}$ in peak $\rm H\alpha$ flux, with a single spectral element. 

All 12 hours were granted to us by the Time Allocation Committee. The observations were broken up into three, four-hour observing blocks to be started October 3, 2018. This was to accommodate the +60$\rm ^o$ declination of the galaxy, resulting in it being at an airmass less than 1.5 for only about four hours per night as seen from Mauna Kea. Unfortunately, uncooperative weather resulted in only one of the three observing blocks being completed and we only acquired four hours of data. 

The standard data reductions were completed on site with the \textsc{orbs}\footnote{Outil de R\'{e}duction Binoculaire pour SITELLE} software package, version 3.4\footnote{https://github.com/thomasorb/orbs} \citep{martin2015orbs}. These include standard astronomical reductions such as bias, dark frame, and flat field corrections, as well as cosmic ray removal. In addition there are reductions specifically required to process IFTS data. \textsc{orbs} completes the combination and alignment of the two interferograms, using background unsaturated stars for reference. It also includes a phase correction to take into account the off-axis configuration of SITELLE as well as the phase solution's wavelength dependence coming from the properties and alignment of the optical elements. Lastly, \textsc{orbs} enables flux and wavelength calibrations, meaning no further data reductions are required once the datacube has been received.

We estimate that the loss of observing time increased our limiting surface brightness from 24.9\,mag/arcsec$^2$ in the R-Band, to 24.3\,mag/arcsec$^2$, or $\rm 6\times10^{-16}\,erg/s/cm^2/\textrm{\AA}$ in peak $\rm H\alpha$ flux per resolution element, which had a substantial impact on our potential to detect LEs. The reduction in depth achieved required extra emphasis be placed on developing a LE search methodology which made the most of the observations that we did receive. The calibrated datacube can be retrieved from the CADC with product ID 2307000p. 


\section{Data Analysis}
\label{sec:Data Analysis}

As can be seen in the top panel of Figure \ref{fig:broad}, SITELLE does not have a filter that completely encompasses an H$\alpha$ line broadened by the velocities expected of a SN. The SN3 filter captures the 6563\,\AA\, rest frame wavelength of the H$\alpha$ line, but due to the filter's 340\,\AA\, bandwidth, broadening by $\gtrsim$4000\,km/s will cause the line to completely fill the bandpass. As the SN3 bandpass covers the range of 6510\,-\,6850\,\AA, it captures only the downslope of the red-shifted emission component of a P-Cygni profile. In our SITELLE data, a highly broadened H$\rm \alpha$ P-Cygni profile that we expect from SNe, would be virtually indistinguishable from negatively sloped continuum emission.

Unfortunately, the blue-shifted absorption component falls between the SN3 and neighbouring C2 filter (5620\,-\,6250\,\AA), meaning we could only observe the emission component of a P-Cygni profile. Such a situation is less than ideal since it prevents us from searching our spectra for full P-Cygni line shapes.

\subsection{A Statistical Methodology for Continuum Fitting}
\label{sec:ContFit}

For all but the hottest stars ($\rm T<10000\,K$), the peak of the SED is situated in, or near to the wavelength range corresponding to the SN3 filter bandpass -- the great majority of stellar continua would appear flat within our data. Therefore, in order to locate potential LE candidates, it would be sufficient to identify regions within our data, where spectra display negatively-sloped continua. This avoids the need for the introduction of extra parameters to constrain, as would be the case if we instead fit for a full P-Cygni profile.

Our continuum fitting methodology relied heavily on the optimized curve fitting framework provided by the \textsc{scipy} Python package\footnote{https://www.scipy.org}, and its least-squares fitting capabilities which can be adapted by the user for a variety of non-linear data models. We also made use of \textsc{orcs}\footnote{Outils de R\'eduction de Cubes Spectraux}, and its backend \textsc{orb} -- the data processing software packages designed specifically to handle SITELLE data \citep{martin2015orbs}. They are publicly available to download\footnote{https://github.com/thomasorb}. \textsc{orb} version 3.4, \textsc{orcs} version 2.4, and \textsc{scipy} version 1.2 were used in this work.

Our datacube consists of 49 frames, corresponding to spectra with 49 spectral points. It is important to note that in all SITELLE datacubes, regardless of the chosen filter, the spectra are sampled over a number of points outside of the bandpass of the filter in order to avoid spectral folding during the Fourier transform in ORBS. With SN3, approximately half of the spectral points are outside of the filter bandpass, irrespective of the spectral resolution. Those spectral points do not contain emission since the filter transmission is null at these wavelengths. In our observations 29 spectral points are outside of the filter bandpass. 

More specifically for the continua fitting technique used in this paper, spectral points over emission line complexes (here H$\alpha$\,+\,[NII]$\lambda$6548, 6583, and [SII]$\lambda$6716, 6731) must be excluded from the fit. NGC 6946 is a low inclination galaxy at $\rm 38^o \pm 2^o$ \citep{boomsma2008hi}. Therefore, there is a velocity component along the line of sight that varies with respect of the North-angle position of the pixels in the FOV. A fit performed with ORCS over the H$\alpha$ line profile determined that the centroid position varies on the order of $\sim$150\,km/s across the galaxy; in good agreement with what was reported by \citet{boomsma2008hi}. Due to our low resolution and spectral point separation of $\sim$18\,\AA, the line complexes' flux do not significantly shift away from the spectral point in which they are located if at rest. This facilitates the exclusion of the emission line spectral points. Eight spectral points (each corresponding to a spectral frame in the datacube), spanning the wavelength ranges 6514.0\,-\,6575.6\,\AA and 6702.8\,-\,6734.7\,\AA, containing the H$\alpha$\,+\,[NII]$\lambda$6548, 6583 and [SII]$\lambda$6716, 6731 complexes respectively were identified and excluded from the fit. 

The remaining 12 frames out of 49, which solely contain continuum emission were therefore used for the final fit. With only 12 data points, there is a risk of over-fitting and possibly inferring the presence of a LE where in reality there is only noise. 

Two related statistical criteria were applied to estimate the significance of a negatively (or positively) sloped continuum over the null-hypothesis (flat continuum): the Akaike Information Criterion \citep{akaike1973maximum, gordon2015regression} and the Bayesian Information Criterion \citep{schwarz1978estimating}, hereafter AIC and BIC respectively. 

When BIC and AIC values are known for a number of competing models, the model which minimizes its information criteria is selected as the most likely to accurately represent the data. In order to test the statistical significance of any continuum slope, a sloped (two-parameter) and flat (one-parameter) model were fit to the extracted spectra. The BIC and AIC values for each model were compared in the following manner:
\begin{align}
\Delta AIC &= AIC_{\textrm{2p}} - AIC_{\textrm{1p}} \\
\Delta BIC &= BIC_{\textrm{2p}} - BIC_{\textrm{1p}} 
\end{align}
A negative value for each criterion indicates that the two-parameter fit is most likely. In particular, values of $\rm \Delta BIC < -2$ and $\rm \Delta AIC < 0$ were our requirements for a statistically significant sloped continuum. If the statistical tests did not meet the above criteria, the continuum was deemed to be consistent with flat.
 
It is known that in SITELLE datacubes as in other imaging instruments, there is a slight variation in the spectral sampling of the filter across the FOV. In addition, SITELLE has a slight variation in spectral resolution across the FOV and an instrumental line shape of a sine cardinal \citep{baril2016commissioning}. Due to these combined effects, careful attention must be taken to the sky subtraction to avoid residual artifacts. Furthermore, due to SITELLE's FOV being nearly identical to the size of NGC 6946 on the sky, there are very few resolution elements in the frame completely free of line emission, further complicating the creation of a sky image.

\begin{figure} 
	\centering
	\includegraphics[width=0.89\columnwidth]{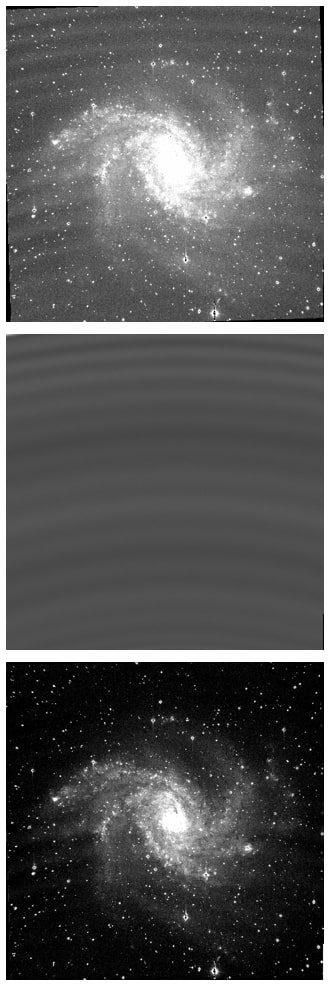}
    \caption{Example of the sky subtraction process. Top: a raw datacube frame. Middle: the extrapolated sky fit for the entire frame. Bottom: the resulting frame after the sky subtraction is completed.}
    \label{fig:skysub}
\end{figure}

We completed our sky correction via a modified image subtraction method. In order to capture the effects of the bandpass, spectral resolution, and sky brightness variation across the FOV, we used a two-dimensional, polynomial weighted, sine and cosine fit to the intensity of pixels with signals less than 1580\,ADU in the deep frame - which contain the least line emission. We then extrapolated the fits from these regions to obtain sky intensity values for all other pixels in the cube, thereby creating sky maps which encapsulate the bandpass and line shape variations. An example of this process is shown in Figure \ref{fig:skysub}.

\subsubsection{Understanding Systematic Biases}
\label{sec:bias}

Due to the depth of our observations, the very brightest of stars are saturated, and display prominent diffraction lines that contaminate the data. These are easily identifiable and a pixel mask was manually constructed to not only mask the point spread functions (PSFs) of the saturated stars themselves, but also the diffraction peaks.

NGC 6946 is classified as a starburst galaxy, and its myriad HII regions, as well as other star forming regions are clearly visible and well studied \citep{schinnerer2006molecular}. These regions are somewhat problematic for our study because it is nearly impossible that we will find a LE buried in the bright emission of such areas. Unfortunately, this means that our methodology contains a bias against LE detection in areas of high emission.

To simultaneously remove the saturated and bright stars, as well as the HII regions from our analysis, a pixel mask was created combining the saturated star mask and a second mask including all pixels with flux >$5000\,\rm ADU$ in the deep image. These masked sources correspond to roughly the brightest 1.5\% of pixels, and includes all background and foreground stars, as well as the cores of HII regions and OB associations. 

Unlike the continuum emission of older stellar populations which appear flat in the SN3 filter, residual emission from young and hot OB type stars will manifest themselves as negative slopes in the SN3 filter, introducing a possible degeneracy between emission from a hot star, or a LE candidate. However, the continuum slope of a hot star on the Rayleigh-Jeans tail is very specific, $\rm (dF_\lambda/d\lambda)/F_\lambda = 4/\lambda$. Any slope caused by a LE would also include H$\alpha$ line emission, and would therefore be steeper than the $\rm 4/\lambda$ expected for hot stars alone. 

A second aid to the resolution of this degeneracy exists in the form of high-spatial resolution photometric observations with the Hubble Space Telescope (HST). NGC 6946 is nearly completely covered by HST observations, which can allow us to better identify regions of the galaxy which may be contaminated by emission from hot stars. We downloaded archival HST images from the ACS/WFC (Advanced Camera for Surveys/Wide Field Channel) from 2016 to construct a mosaic image of NGC 6946\footnote{CADC observation IDs jd9g01020, 02020, 03010}, clearly showing the locations of the brightest contaminating objects. We can thus cross-reference the locations of LE candidates discovered through our slope search method, with positions of stars in the archival HST data.

When running the continuum fitting procedure, pixels were binned on a 7x7 scale in order to increase the S/N. Binning over 49 spectral elements decreases our survey's surface brightness limit to $\rm 2\times10^{-17}\,erg/s/cm^2/arcsec^2/\textrm{\AA}$, potentially enabling the re-detection of the SN 2002hh LE. Integrated spectra were extracted over the 7x7 binned-pixel datacube and then fitted. This binning also corresponds to approximately the size of SITELLE's worse PSF measured on the raw frames of the datacube. Therefore, we are not losing much spatial information from the integration but are increasing the S/N of the spectra. During the binning, pixels which were to be masked using the method explained above were not included in the integrated spectrum of each region. Additionally, if more than 50\% of a region was masked, the region as a whole was rejected. Out of 64,009 binned pixels, a total of 2,811, or 4.4\%, were masked.


\section{Results}
\label{sec:Results}

\subsection{Searching for Light Echoes from the Ten Supernovae}

There are no detailed dust maps of NGC 6946 to guide our search for LEs, and as such we had to search the entire area over which it would be possible to find a LE from each SN. From the known geometry of scattered LEs, there is a maximum projected area around the SN epicentre within which it is theoretically possible for a LE to be found which, of course, wholly depends on the existence and favourable positioning of scattering dust within this volume.

The radius of the circular area is determined through use of the LE equation \citep{couderc1939aureoles}. 

\begin{equation}
\rho^2 = 2zct + (ct)^2
\label{eq:LEequ}
\end{equation}

\noindent Here, $z$ is the distance along the line of sight from the SN to scattering dust, $\rho$ is a projected distance in the plane of the sky, $t$ is the time since the SN eruption, and $c$ the speed of light.

In addition to the age of the SNe, and the 7.72\,Mpc distance to NGC 6946, an assumption needed to be made about the dust height above the galactic plane, $z$, in order to calculate the projected radius, $\rho$. The thin disk contains most of the galactic dust, so we assumed a thin disk scale height of 0.25\,kpc for NGC 6946 -- comparable to that of the MW \citep{carroll2017introduction} representing the maximum height above the galactic plane within which we could reasonably expect the presence of dust in a significant amount.

\begin{figure} 
	\centering
	\includegraphics[width=\columnwidth]{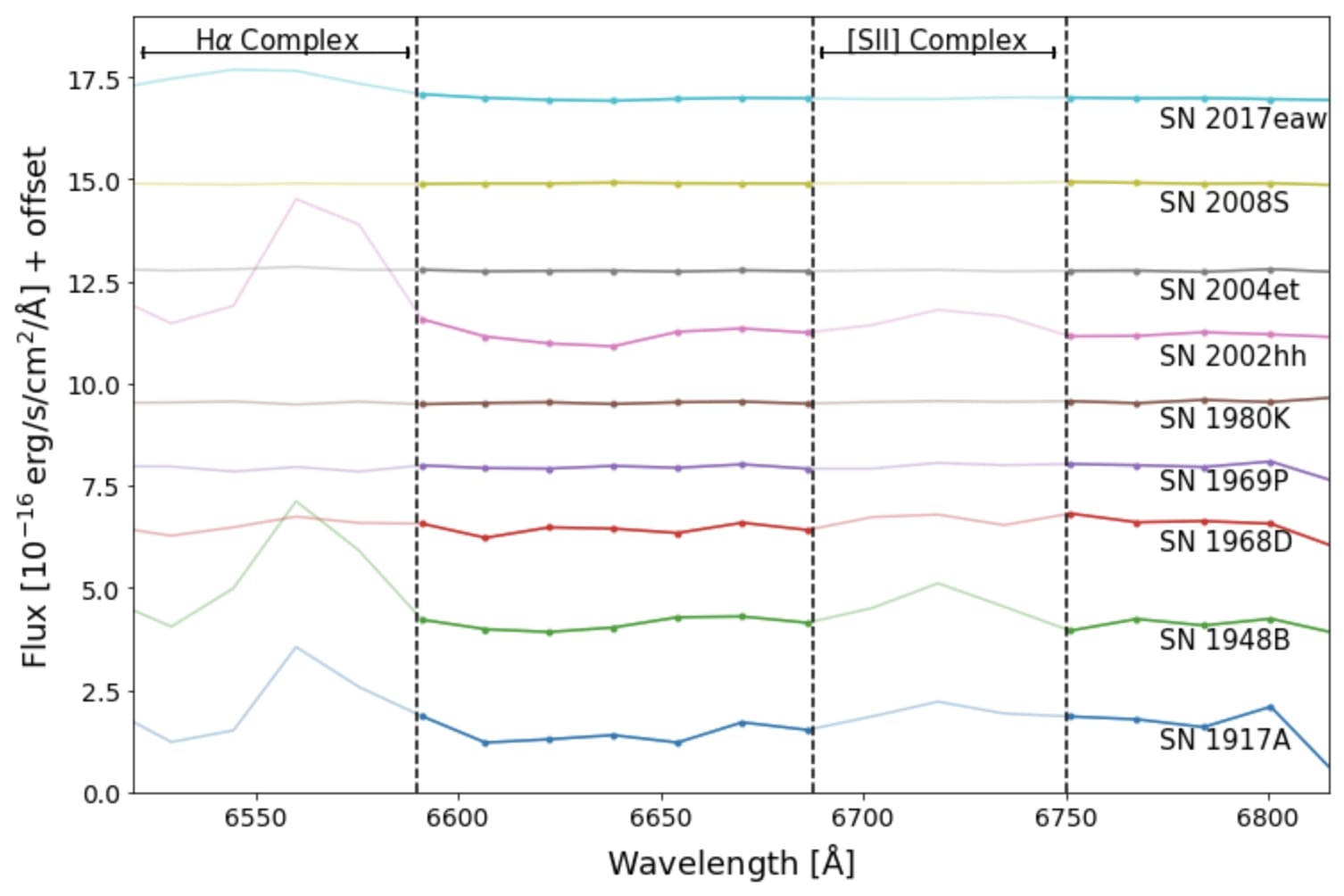}
    \caption{Extracted spectra from the ten SN regions (SN 1939C is excluded due to contamination from a foreground star). The faded regions represent the wavelength ranges excluded from the fit due to the presence of emission lines. The points on the spectra denote the 12 remaining frames used in our fits. Each spectrum is given an arbitrary offset for visual clarity.}
    \label{fig:SNSpec}
\end{figure}

With these assumptions, we determined the projected LE search area for each of the ten known SNe. An integrated spectrum was extracted from each region for further examination for evidence of any LE candidates. We fit the continuum in each of the extracted spectra to determine whether any of these regions displayed the characteristic negative slope that we anticipate will indicate the presence of LEs. The region parameters, and results of the continuum fitting are summarized in Table \ref{tab:LEindivparam}, and the individual spectra are displayed in Figure \ref{fig:SNSpec}.

\begin{table}
\centering
\caption{Summary of the parameters, and continuum fitting results from the initial search regions around the ten recorded SNe}
\begin{tabular}{|c|c|c|c|c|}
\hline
\hline
\multirow{2}{*}{Name} & Radius  & Slope & \multirow{2}{*}{$\Delta$BIC} & \multirow{2}{*}{$\Delta$AIC} \\
& (arcsec) & $\rm (erg/s/cm^2/\textrm{\AA}/\textrm{\AA})$ & & \\
\hline
1917A & 3.43 & 7(12)$\times10^{21}$ & 2.5 & 8.0 \\ 
1939C & 3.01 & -1.7(7.2)$\times10^{-19}$ & 2.4 & 7.9 \\
1948B & 2.82 & -1.4(1.0)$\times10^{-19}$ & 0.3 & 5.8 \\
1968D & 2.37 & 3(76)$\times10^{-21}$ & 2.5 & 7.9 \\
1969P & 2.35 & -1.4(4.7)$\times10^{-20}$ & 2.4 & 7.9 \\
1980K & 2.06 & 2(1)$\times10^{-21}$ & -0.8 & 4.8 \\
2002hh & 1.33 & -9.7(8.8)$\times10^{-20}$ & 1.1 & 6.6 \\
2004et & 1.24 & -7(8)$\times10^{-21}$ & 1.5 & 7.1 \\
2008S & 1.05 & -1(6)$\times10^{-21}$ & 2.4 & 8.0 \\
2017eaw & 0.33 & -1.3(6.5)$\times10^{-19}$ & -1.6 & 3.9 \\
\hline
\label{tab:LEindivparam}
\end{tabular}
\end{table} 

There was nothing immediately obvious in any of the spectra that suggest the presence of a LE. The continuum fitting results also confirm the lack of clear evidence for SN LEs in any of these regions. None of the ten regions meet the criteria of having a continuum slope that can only be negative within error, as well as values for the statistical criteria of $\Delta$BIC < -2 and $\Delta$AIC < 0. In other words, all of the continua are consistent with being flat. 

It is somewhat odd that we do not detect emission, or evidence of LEs from SN 1980K. Late-time emission was observed photometrically at a level of $\rm \lesssim 10^{-15}\,erg/s/cm^2$ in 2010 \citep{milisavljevic2012late}. It was also observed spetroscopically by \citet{long2019new}. Although the late-phase emission is predominately line emission, the spectrum presented by \citet{long2019new} displays a boxy H$\rm \alpha$ line typical of a SN with strong circumstellar material interactions \citep{morozova2017unifying}. Thus, such boxy emission may be missed by a search method centered around identifying strong slopes. 

In the interest of completion, we also examined the region around the failed SN candidate identified in NGC 6946 by \citet{gerke2015search} for evidence of LEs. Our analysis once again indicated a flat spectrum, showing no evidence for LEs. This is perhaps unsurprising as the most recent photometric observations of this target from December 2015 find a magnitude of $\sim$24.8 in the R-band \citep{adams2017search}. Any remaining optical emission has likely faded well below the limit of our survey in the intervening time.

\subsection{Searching for Light Echoes from Historical Supernovae}

\begin{figure*} 
	\centering
	\includegraphics[width=0.9\textwidth]{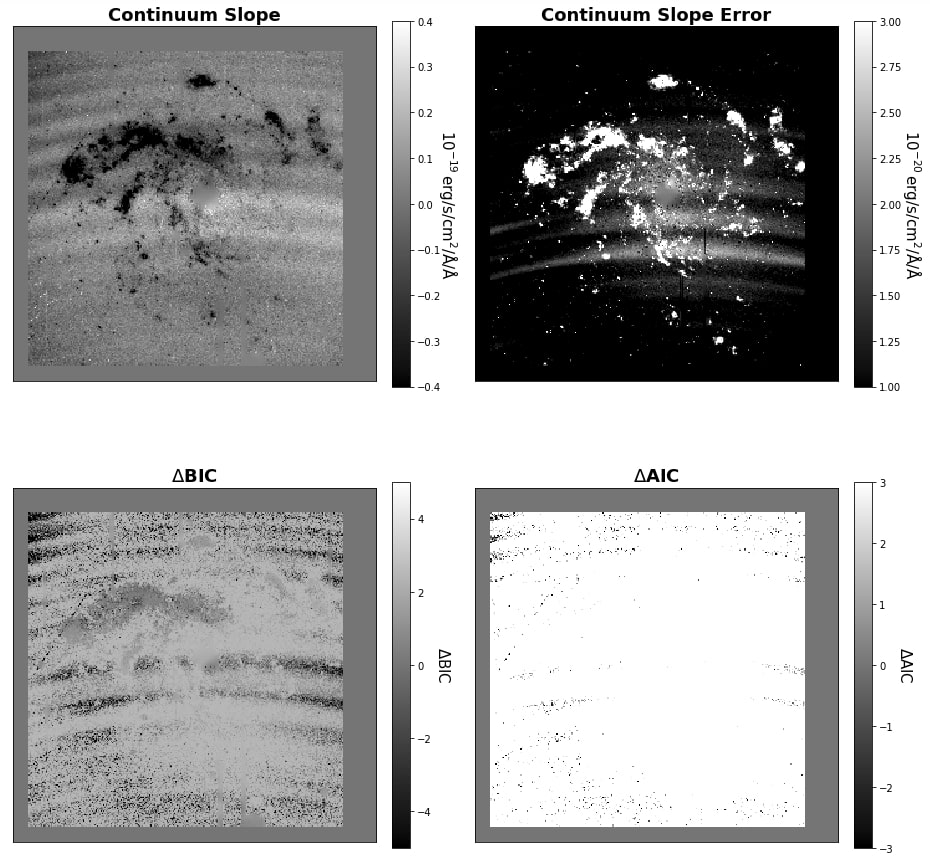}
    \caption{Maps of the continuum slope, error in continuum slope, $\Delta$BIC, and $\Delta$AIC parameters for each 7$\times$7 binned pixel in the frame. The masked areas have been filled with an interpolation of the surrounding pixels.}
    \label{fig:4-newres}
\end{figure*}

As already stated, it is entirely possible that within NGC 6946 there are a set of detectable LEs from SNe predating the discovery of SN 1917A. Searching for negatively-sloped continua throughout the rest of the galaxy then serves a dual purpose: it could yield LE candidates from such historical SNe, and even if none are found it will serve as a calibration for the false-positive detection rate for our continuum-fitting search strategy. The final maps of continuum slope, error in continuum slope, $\Delta$BIC, and $\Delta$AIC are shown in Figure \ref{fig:4-newres}.

All four maps are fairly uniform -- the major features being the ghostly image of the galaxy, as well as some low-level, background arcs which are residual artifacts of the aforementioned variations in the instrumental line shape and the spectral sampling of the filter. This uniformity is to be expected; the vast majority of pixels should hold only stellar emission, yielding flat continua with relatively small errors. This is precisely what is seen. The masked regions have been filled with an interpolation of the surrounding pixels, which adds to the smooth nature of the images.

Some features of the galaxy remain visible in the slope map -- namely the star-forming ISM is traced. It is unclear why some of these features appear so strongly in this map - possibly buried within the ISM are more hot stars or the continuum scattered from the neighbor massive stars impart a slight negative slope onto the continuum emission in these regions. However it is notable that these same portions of the spiral arms have also the largest values in the error map. So even though some regions of the ISM appear to have largely sloped continua, these same regions also have the largest uncertainties; preventing the measurement of a clear negative slope. This is further emphasized by the $\Delta$BIC, and $\Delta$AIC maps, in which almost no structures tracing the galaxy are present.

Despite our best efforts to minimize them, artifacts of the filter bandpass, and instrumental line shape variations are still present across all four of our maps. The effects appear to be relatively low-level in the slope and error maps, however in the $\rm \Delta AIC$ map, it can be clearly seen that the vast majority of pixels with large negative $\rm \Delta AIC$ values directly trace these artifacts.

The goal of our analysis was to locate potential LE candidates via regions with statistically-significant negative slopes. This required a binned pixel to have a consistent negative slope within its error, as well as $\Delta$BIC < -2 and $\Delta$AIC < 0 to indicate that the two-parameter fit is significant. Out of 61198 unmasked, binned pixels, only 149 met these criteria. However, upon subsequent visual inspection of these regions, it became clear that all but one of them clearly traced the instrumental artifacts. The other 148 potential detections were therefore discarded as false positives, and we focused on the sole region left which met our LE criteria - located at $\rm \alpha = 20:34:40.218$, $\rm \delta = +60:04:39.49$ (J2000).

In order to attempt to determine whether the sloped continuum in this region was indeed caused by a LE, we first cross-referenced its position with our archival HST data. The region of interest was free from any bright stars, and thus unlikely to be contaminated by residual emission from a hot star. We then proceeded to fit a model P-Cygni profile to the extracted spectrum. Although the best fitting peak $\rm H\alpha$ flux is a reasonable value of $\rm 2\times10^{-17}\,erg/s/cm^2/\textrm{\AA}$, the broadening extends to clearly nonsensical values in the two-dimensional posterior distribution with significant probability. In fact, the broadening does not fall below $\rm 10000\,km/s$ at even the $\rm 5\sigma$ level, indicating that it is very unlikely that the negative slope in this region was indeed caused by the presence of a LE.

Our continuum fitting method presents a false-positive rate of approximately 0.24$\%$. This takes into account the 148 regions which were clearly biased by the instrumental artifacts, as well as the one remaining region which we concluded did not contain a LE. Further applications of this method at a higher spectral resolution where the data is not effected by filter bandpass and line shape variations would likely reduce, or even eliminate the number of regions biased by artifacts, and obtain a truer estimate of the false positive rate of this method. However, until such analysis is possible, 0.24$\%$ serves as an upper limit on that rate. 

\subsection{A Flux Limit for Light Echo Detection}
\label{sec:FLFLE}

Given that we did not manage to detect any LEs, a firm limit in age and surface brightness leading to detectable LEs cannot be established. However, we can set an upper limit in peak H$\alpha$ flux with the data we have available. Model P-Cygni profiles were inserted at random locations into the galaxy, and we attempted to detect them with our continuum slope method. H$\alpha$ P-Cygni line profiles, broadened by 200\,\AA, and with peak flux ranging from $\rm 5\times 10^{-18}$ to $\rm 2.0\times 10^{-16}\, erg/s/cm^2/\textrm{\AA}$ were tried. At each peak flux, 50 P-Cygni profiles were inserted into random locations in the galaxy. Our continuum fitting methodology was applied to each of the 50 regions to see how many times the continuum fit indicated the presence of a LE candidate. 

We consider our detection limit to be the point at which we recover the P-Cygni model 50\% of the time. After the aggregation of results over ten trials, we determined a limit in peak flux of $\rm 3\times10^{-17}\,erg/s/cm^2/\textrm{\AA}$, or $\rm 1\times10^{-15}\,erg/s/cm^2/arcsec^2$ in surface brightness flux. Therefore, any LEs present within NGC 6946 must be at least this faint in H$\alpha$ emission or else we would have detected them in this work. The 2015 peak flux of the SN 2002hh LE, falls exactly on this limit. However, if we assume that the LE faded at a constant rate from 2004 to 2015, and continued to fade at the same rate in the three years leading up to our observations, the LE would have a peak $\rm H\alpha$ flux of only $\rm 2.2\times10^{-18}\,erg/s/cm^2/\textrm{\AA}$, which falls well below our detection limit.

We believe that the analytical approach described in this section would benefit any future LE searches in NGC 6946. Such studies would need to reach at least the level of our detection limit in H$\alpha$ in order to undertake a meaningful LE survey. However, it should be noted that approximately an order of magnitude in additional sensitivity would likely be necessary to establish a statistical distribution of LEs within NGC 6946.


\section{Discussion \& Conclusions}
\label{sec:Conclusions}

Unfortunately, our survey yielded no LEs from the ten known SNe, or from any SNe prior to 1917. Any detections were likely confounded in part by the lack of observation depth, we had proposed for and were granted a total of 12 hours of time. Nevertheless, we established an upper limit on LE H$\alpha$ surface brightness of $\rm 1\times10^{-15}\,erg/s/cm^2/arcsec^2$ in this survey. It is in retrospect unlikely that even with the full 12 hours of observation time, we would have been able to establish the statistical sample of LEs for which we had hoped.

Another difficulty, although of secondary importance, encountered in the data was the result of instrumental artifacts caused in part by our operation of SITELLE at such a low spectral resolution. SITELLE is primarily used for moderate spectral resolution studies (R between 1000 and 5000), and we have obtained the first ever observations with SITELLE at such a low spectral resolution. As such, our data is the only set available in which these artifacts remain, and we were able to make the best out of it by adjusting the method for the sky subtraction. In this way, our observations, and attempts at correction of the artifacts are meritorious for the development of SITELLE itself. 

The width of the SN3 filter was both a help, and a hinderance throughout this study. Although its narrow bandpass allowed us to approximate a P-Cygni profile as a single negative slope, fitting the entire P-Cygni profile would lead to a more unequivocal detection. Unfortunately, SITELLE has no filter which captures the absorption component of the P-Cygni. A wider SITELLE H$\alpha$ filter would allow for detection, and confirmation of a candidate LE in a single observation. Short of that, the same length of observations in the neighbouring SITELLE filter, C2, would provide better characterization of the stellar continuum. C2 (5590\,-\,6240\,\AA) falls outside the range of the highly-broadened H$\alpha$ line, and deviations in the continuum from C2 to SN3 would provide stronger evidence that negatively-sloped continua are due to a LE. 

In striving for the deepest possible observations, we likely also sacrificed too much in spectral resolution. Our chosen resolution resulted in only 12 frames encompassing the entirety of the continuum emission in the SN3 filter. The prominent emission lines, H$\alpha$, [NII]$\rm \lambda$6548, 6584, and [SII]$\rm \lambda$6717, 6731 were combined into two broad, unresolved groups, and so could not be individually subtracted. Our low spectral resolution effectively smeared the emission lines over a wider wavelength range. For example, the peaks of the [SII] doublet are separated by only 14\,\AA, however in our observations, emission from the doublet is seen in spectral points spanning 6680 to 6760\,\AA. The 14\,\AA gap has been spread to $\sim$80\,\AA. 

Increasing the spectral resolution would not change the fraction of frames lying outside the filter bandpass, but it would change how much the emission lines are smeared into neighbouring frames. SITELLE's sine cardinal line shapes also come with additional challenges. Very strong emission lines, such as H$\alpha$, will have prominent side-lobes in IFTS data. The strongest emission lines will always spill into neighbouring continuum frames because of these side-lobes, but this effect is much weaker for the fainter [SII] lines. Fitting the narrow lines emission using the sinc function and subtracting that model before measuring the continuum might be an avenue for future work.

With these considerations, a spectral resolution of at least R\,=\,600 would likely be ideal for a future SITELLE survey. This would provide enough resolution to at least differentiate the individual emission lines in the two groups, and increase the number of spectral points to be used for the continua fitting.

Moreover, R\,=\,600 is not so high of a spectral resolution such that the S/N we could hope to reach is detrimentally effected. For sky limited, or bright sources there is a reduction in continuum S/N by a factor of $\rm 1/\sqrt{2}$ for each factor of two increase in spectral elements. By increasing our spectral resolution from R\,=\,300 to R\,=\,600, we are effectively dividing each spectral element in two, and suffering the equivalent penalty in the S/N of the continuum. For example, based on outputs of the SITELLE exposure time calculator\footnote{http://etc.cfht.hawaii.edu/sit/}, with our initial observing parameters we would see a LE at the brightness of the \citet{andrews2015late} detection of the SN 2002hh LE with a S/N of $\sim$1.5. With a spectral resolution of R\,=\,600 and the same total length of observations, we would still reach a S/N of $\sim$1.1, but have more than double the number of frames to which a P-Cygni profile could be fit. Increasing the total observation time by an additional three hours would allow us to reach the same S/N as with R\,=\,300. 

This work has provided a methodology to enable LE candidate identification in single-observation spectroscopic data, however the ideal combination of instrument and target have yet to be identified. We have taken steps to resolve the largest remaining problem in the study of LEs - the development and observation of a program to identify a statistical sample of LEs, and set firm detection limits. A revisiting of this work with longer observations and a higher spectral resolution may yet prove the most fruitful route to this goal. 

\section*{Acknowledgements \& Data Availability}

The authors would like to thank the anonymous referee for helpful comments regarding this manuscript. M.R. acknowledges support from the Natural Sciences and Engineering Research Council of Canada for support towards his M.Sc studies. 

The data underlying this article can be retrieved from the Canadian Astronomy Data Center at http://www.cadc-ccda.hia-iha.nrc-cnrc.gc.ca/en/ with product ID 2307000p.


\bibliographystyle{mnras}
\bibliography{Bibliography} 

\bsp	
\label{lastpage}
\end{document}